\documentclass[aps,preprint,a4paper,nofootinbib,superscriptaddress]{revtex4}
\usepackage[utf8]{inputenc} 
\usepackage[T1]{fontenc} 
\usepackage[dvipsnames]{xcolor}
\usepackage{epsf, epsfig, graphicx,wrapfig,subfigure,float,url,ulem,enumitem,footnotebackref}
\usepackage{amsmath, amssymb, amsthm, physics, braket,mathrsfs,nccbbb,cancel,mathtools}
\usepackage[toc,page]{appendix}
\usepackage[hang, flushmargin]{footmisc}
\usepackage[mathscr]{eucal}
\usepackage{color}

\newcommand{\be}{\begin{equation}}
\newcommand{\ee}{\end{equation}}
\newcommand*\Eval[3]{\left.#1\right\rvert_{#2}^{#3}}

\begin{document}

\title{Entanglement Islands from Holographic Thermalization of Rotating Charged Black Hole}
\author{Po-Chun Sun}
\email{jim.pochun.sun@gmail.com}
\affiliation{Department of Physics, National Central University, Chungli 32001, Taiwan.}
\date{August 21, 2021}

\begin{abstract}
We study the time evolution of the entanglement entropy of Hawking radiation in the $(n+1)-$dimensional Kerr-Newman black hole evaporation by the holographic approach that considering the $(n+1)-$dimensional AdS eternal black brane coupled to the auxiliary CFT reservoir is in the Hartle-Hawking state. The CFT reservoir itself has a holographic dual, the $(n+2)-$dimensional bulk geometry, and the original $(n+1)-$dimensional AdS-black brane is embedded into such bulk manifold, which is precisely Randall–Sundrum model. 
  
  According to the island rule~\cite{Almheiri:2020}, the entanglement entropy in semi-classical gravity can be divided into two parts, one is due to the quantum effects, which can be obtained by Ryu–Takayanagi conjecture. Another is the gravitational part, which is equal to the area of the quantum extremal surface divided by four times the Newton's constant. We show that the entanglement growth in our holographic system is linear in late times. After Page time, the system reaches saturation since the entanglement islands appear. In this paper, we will emphasize how black hole rotation affects entanglement entropy in general dimensional spacetime.
\end{abstract}

\maketitle


\section{Introduction}
Quantum mechanics is a fundamental theory that tremendous success at the microscopic level. On the other hand, general relativity is a useful theory of gravitation that passes many experiment tests. Nevertheless, from the combination of two theories, the black hole information paradox is a big conundrum for decades. Let’s illustrate that with an explicit example. Starting from a star in pure state collapses to form a black hole, and then some entangled pairs are created near the event horizon such that those particles go out the black hole entangled with the particles inside the black hole. It is turn out to be a mixed state when the black hole completely evaporates, which contradicts the time evolution symmetry (the whole system is unitary) according to quantum mechanics.

The papers~\cite{Penington:2020,Almheiri:202003} showed that the entanglement entropy increases at early times in the system, the black hole is evaporated by adding an auxiliary quantum thermal bath (the black hole with a thermal bath in the Hartle-Hawking state), and the entanglement entropy decreases in late times owing to the appearance of the quantum extremal island, which leads to the reconstruction of the entanglement wedges.~\cite{Almheiri:2020} provided a new way to think about such $(n+1)-$dimensional gravitational system add quantum matters couple to a thermal bath. There are three different alternative descriptions: 
\begin{enumerate}
    \item Bulk Perspective: describing the quantum matters as more than one dimension gravitational dual (bulk manifold). The original $n-$dimensional gravitational system lives on the dynamical brane (Planck brane), and the Planck brane is embedded in the bulk manifold.
    \item Brane Perspective: the ordinary picture which we have described.
    \item Boundary Perspective: describing the gravitational part as less than one dimension QFT dual.
\end{enumerate}
\begin{figure}
\centering
{\setlength{\fboxsep}{0pt}\setlength{\fboxrule}{1pt}\fbox{\includegraphics[width=0.28\textwidth]{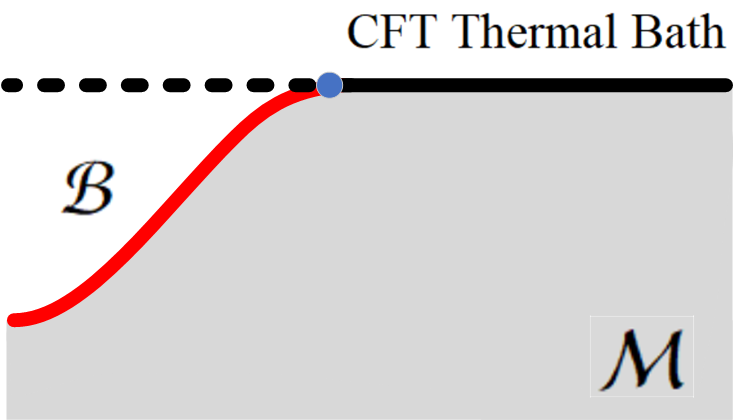}}}
{\setlength{\fboxsep}{0pt}\setlength{\fboxrule}{1pt}\fbox{\includegraphics[width=0.36\textwidth]{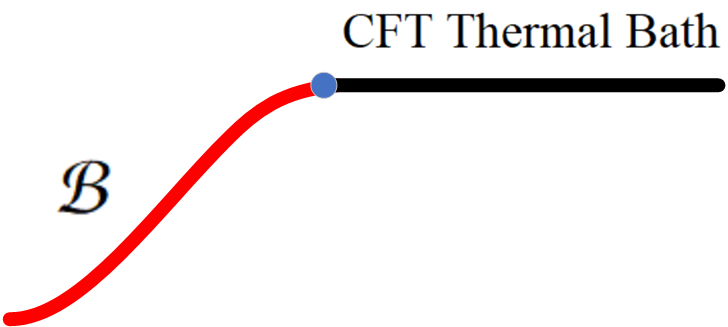}}}
{\setlength{\fboxsep}{0pt}\setlength{\fboxrule}{1pt}\fbox{\includegraphics[width=0.265\textwidth]{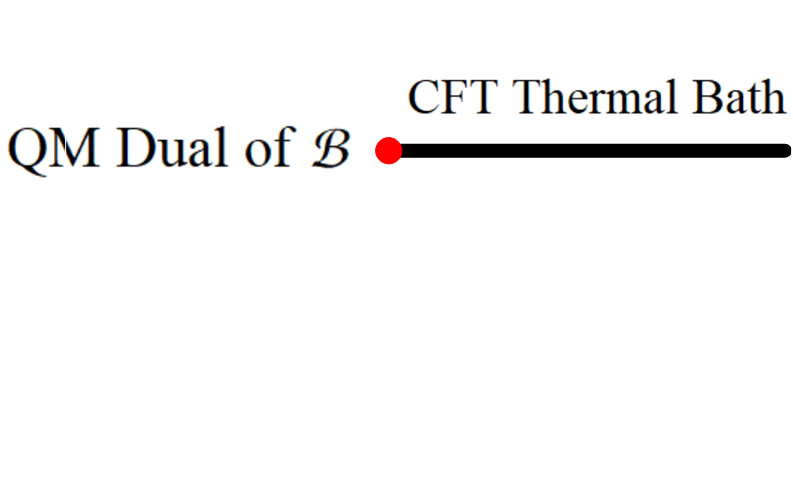}}}
\caption{The $\mathcal{M}$ is the $(n+2)-$dimensional bulk manifold and the $\mathcal{B}$ is the $(n+1)-$dimensional Plank brane. {\it Left}: bulk perspective; {\it Middle}: brane perspective; {\it Right}: boundary perspective.}
\end{figure}
Apart from that, let's say $\Upsilon$ is the radiation region in the quantum reservoir, then, according to the island rule is proposed by~\cite{Almheiri:2020}, the von Neumann entropy in semi-classical approximation can be computed by
\be\label{island-rule}
\mathcal{S}_\Upsilon=Min\left\{\underset{\mathcal{I}}{Ext}\left[\mathcal{S}^{\text{(eff)}}_{QFT}(\Upsilon\cup\mathcal{I})+\frac{\mathcal{A}_{\partial \mathcal{I}}}{4G^{(brane)}_N}\right]\right\}\,.
\ee
where the corresponding reduced density $\rho^{\text{(eff)}}_{\Upsilon\cup\mathcal{I}}$ in semi-classical limit is different from the reduced density $\rho_{\Upsilon\cup\mathcal{I}}$ in the exact quantum theory. Hence, the metric we consider in this paper is always classical gravity. Using the RT formula \cite{Ryu:2006bv,Ryu:2006ef}, we can calculate the entanglement entropy in QFT by $\mathcal{S}^{\text{(eff)}}_{QFT}(\Upsilon\cup\mathcal{I})=\frac{\mathcal{A}_\Gamma}{4G^{(bulk)}_N}$
where $\Gamma$ is the RT surface that homologous to $\Upsilon$. There is an explicit computation of time evolution of entanglement entropy and quantum extremal island to resolve the information paradox for two-dimensional AdS black hole in equilibrium (eternal black hole) with a thermal bath in~\cite{Almheiri:2019,Geng:202107039}. The papers \cite{Geng:2020,Geng:202112,Almheiri:202007,Ling:202010,H.-Z.Chen:202006,H.-Z.Chen:202010,Alishahiha:202108715} demonstrated that, even in the higher dimension, the resolution is worked. Nevertheless,~\cite{Bhattacharya:202111870,Bhattacharya:202104134,Bhattacharya:202115852} looked at the aspects of Page curve from the point of view of purification and complexity.~\cite{Geng:202103390} pointed out an inconsistency of entanglement islands in AdS and showed that the resolution of the inconsistency is massive gravitons. Here we need to emphasize that although there is a new prediction that quantum extremal surface appears due to holography, we do not know how Hawking particles bring out the information, what the mechanism is such that the quantum extremal surface looks like a screen. Also, there is an ambiguity how radiation regions increase.

In this paper, we generalize the previous work from \cite{Ling:202010} to rotating black hole. To
simulate the rotating charged black hole in real world, we employ the time evolution of von Neumann entropy in the system, the $(n+1)-$dimensional AdS-Kerr-Newman black hole, with the planar horizon, in equilibrium with a  CFT bath in the Hartle-Hawking state. We analytically calculate the fine-grained entropy in the case that the brane with weak tension limit, which means its backreaction to the bulk geometry is negligible. Before Page time, there is no island $\mathcal{I}=\{\emptyset\}$, the RT surface $\Gamma$ across the bulk horizon. Hence, $\Gamma$ is equal to the wormhole path (Einstein–Rosen bridges). The entanglement entropy increase because the area of the wormhole increase with time~\cite{Carmi:2018}. Particularly, in late times, the entanglement entropy grows linearly because of the stretching of space inside the horizon~\cite{Hartman:2013}. There is no intersection between the entanglement wedges of Hawking radiation and Planck branes until Page time. After Page time, the system reaches saturation and $\mathcal{I}\neq\{\emptyset\}$, the RT surface $\Gamma$ across the Planck brane, the RT surface $\Gamma$ is equivalent to the minimal surface defect in which given strip in $(n-1)-$dimensional spatial region. Due to the reconstruction of the entanglement wedges, there is an intersection between entanglement wedges of Hawking radiation and Planck branes. On top of that, the saturation entropy in the weak tension limit is closed to Bekenstein-Hawking entropy.

The organization of this paper as follows. In section~\ref{sec2}, we build up our doubly-holographic system and illustrate how to embed the Planck brane into the bulk manifold. In section~\ref{sec3}, we write down the general formula for the time evolution of wormhole area, evaluating entanglement velocity at late times. Furthermore, we calculate the entanglement entropy when quantum extremal islands appear in section~\ref{sec4}. In the end, we summarize some important results and discuss some physics meaning in section~\ref{sec5}.
\section{Holography Setup}\label{sec2}
In the paper \cite{Ryu:2006bv}, the framework is underline AdS/CFT, which is that the pure quantum mechanical system dual to the pure gravitational system. Now we study the gravitational system with the quantum matter, and the quantum part has the gravitational dual. We do not know what the precise density matrix $\rho$ is in such quantum gravity, and it is doable in semi-classical gravity because the metric is still classical. Although it may not be a good approximation, hope it gives some clues about quantum gravity and the resolution of information paradox.
\subsection{Review AdS/BCFT}
We consider the AdS/BCFT correspondence~\cite{Takayanagi:2011,Fujita:2011}, denoting $(\mathcal{M},\mathfrak{g})$ as the $(n+2)-$dimensional bulk manifold and $(\mathcal{B},\mathfrak{h})$ as the $(n+1)-$dimensional brane manifold in the bulk, the action can be written as 
\be
I=I_\mathcal{M}+I_\mathcal{B}
\ee
where
\begin{align}\label{m-action}
    &I_\mathcal{M}\equiv \frac{1}{16\pi G^{(n+2)}_N}\int_{\mathcal{M}}d^{n+2}x\sqrt{\abs{\mathfrak{g}}}\left(R-2\Lambda\right)-\int_{\mathcal{M}}d^{n+2}x\sqrt{\abs{\mathfrak{g}}}\frac{\mathcal{F}^2}{4}\,,\\
    &I_\mathcal{B}\equiv  \frac{1}{8\pi G^{(n+2)}_N}\int_{\mathcal{B}}d^{n+1}x\sqrt{\abs{\mathfrak{h}}}\left(K-\mathcal{T}_{\mathcal{B}}\right),~~~K=\mathfrak{h}^{ij}K_{ij}\,.
\end{align}
The quantity $R$ is the Ricci scalar, $\Lambda=-\frac{n(n+1)}{2 L^2}$ is the cosmological constant with the AdS$_{n+2}$ radius $L$, $\mathcal{F}_{\mu\nu}$ is the electromagnetic tensor, $K_{ij}$ is the extrinsic curvature and $\mathcal{T}_{\mathcal{B}}$ is the tension on the brane $\mathcal{B}$.
By varying the bulk action (\ref{m-action}) with respect to bulk metric $\mathfrak{g}_{\mu\nu}$ and the gauge vector of electromagnetic tensor $\mathcal{A}_{\mu}$\,, we get the equation of motion on $\mathcal{M}$ is
\begin{align}\label{eom_m}
    &R_{\mu\nu}+\frac{n+1}{L^2}\mathfrak{g}_{\mu\nu}=8\pi G^{(n+2)}_N\left(T_{\mu\nu}-\frac{T}{n}\mathfrak{g}_{\mu\nu}\right)
\end{align}
where 
\be
T_{\mu\nu}=\mathcal{F}_{\mu\alpha}{\mathcal{F}_{\nu}}^\alpha-\frac{\mathcal{F}^2}{4}\mathfrak{g}_{\mu\nu}\,,~~~~~\nabla_{\mu}\mathcal{F}^{\mu\nu}=0
\ee
and $T=\mathfrak{g}^{\mu\nu}T_{\mu\nu}$. By imposing the Neumann boundary condition on $\mathcal{B}$ ($i.e.$ $\delta\mathfrak{h}_{ij}\neq 0$), we obtain the equation of motion on $\mathcal{B}$~\cite{Chu:2018} is
\begin{align}
    &K_{ij}-K\mathfrak{h}_{ij}+\mathcal{T}_{\mathcal{B}}\mathfrak{h}_{ij}=0\label{eom-on-b}
\end{align}
with the constraint
\be
n^{\mu}\mathcal{F}_{\mu\nu}{P^{\nu}}_j=0\label{em-constraint}
\ee
where $n^\mu$ is the unit vector normal to brane and ${P^{\nu}}_j$ is the projection tensor.
\subsection{Black Brane Thermalization}
Recall the expectation of an observable in a thermal state (grand canonical ensemble
) can be calculated by $	{\left \langle \mathcal{O} \right \rangle}_T=\frac{1}{Z}\Tr \left(\mathcal{O} e^{-\beta (H-\mu \mathcal{Q}+\Omega \mathcal{J})}\right)=\Tr \left(\mathcal{O} \rho_T\right)$ in which $\rho_T=\frac{1}{Z}\sum_{n}e^{-\beta (H-\mu \mathcal{Q}+\Omega \mathcal{J}) }\ket{\psi_n}\bra{\psi_n}$ is the density matrix and $Z$ is the partition function. There is another idea to think about the QFT at finite temperature~\cite{Nishioka:2018}. Consider the total Hilbert space can be written as two copies of the system $H_{tot}=H_L\otimes H_R$ and prepare a special entangled state, the thermofield double (TFD) state
\be\label{thermofield}
\ket{\Psi}=\frac{1}{\sqrt{Z}}\sum_{n}e^{-\beta (H-\mu \mathcal{Q}+\Omega \mathcal{J})/2}\ket{\psi_n}_L\otimes\ket{\psi_n}_R\,.
\ee
Notice that the state (\ref{thermofield}) is invariant under $H_R-H_L$. We see that the reduced density matrix for subsystem $H_R$ is $\rho_R=\Tr_L \left(\ket{\Psi}\bra{\Psi}\right)=\frac{1}{Z}\sum_{n}e^{-\beta (H-\mu \mathcal{Q}+\Omega \mathcal{J}) }\ket{\psi_n}_{R} \prescript{}{R}{\bra{\psi_n}}$, which is the same as the thermal density matrix $\rho_T$. Hence we can interpret the emergence of temperature owing to the ignorance of the subsystem $H_L$.

For bulk perspective, we take the $\mathbb{Z}_2$ quotient~\cite{H.-Z.Chen:202006} across the brane in the bulk manifold $\mathcal{M}$.
\begin{figure}
\centering
\includegraphics[width=0.7\textwidth]{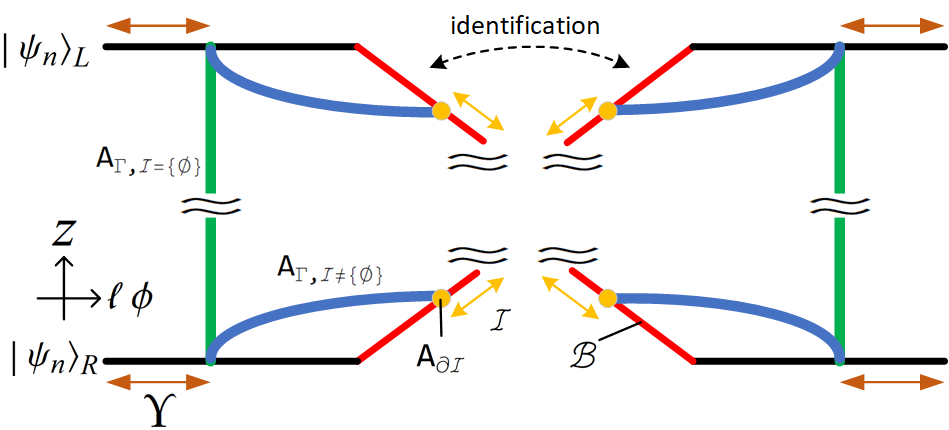}
\caption{The black line is the CFT reservoir; the red line represents the Plank brane $\mathcal{B}$ where the islands $\mathcal{I}\subseteq\mathcal{B}$ appear after Page time. The yellow dots are the quantum extremal surface which are the boundary of the islands $\partial\mathcal{I}=\mathcal{B}\cap\Gamma$. Before Page time, the RT surface $\Gamma$ is the green line. After Page time, the RT surface transit from the green line to the blue line.}
\label{phae}
\end{figure}
To evaluate the von Neumann entropy of the Radiation region, according to the island rule (\ref{island-rule}), we need to compare the no island phase and island phase, and then choose the minimal one
\be\label{evap-ee}
\mathcal{S}_\Upsilon=Min\left\{\frac{\mathcal{A}_{\Gamma,\mathcal{I}=\{\emptyset\}}}{4G^{(\mathcal{M})}_N},\,\underset{\mathcal{I}\neq\{\emptyset\}}{Ext}\left[\frac{\mathcal{A}_\Gamma}{4G^{(\mathcal{M})}_N}+\frac{\mathcal{A}_{\partial \mathcal{I}}}{4G^{(\mathcal{B})}_N}\right]\right\}\,.
\ee

Our consideration of the system is the following: for $\mathfrak{t}<0$, the CFT reservoir and the AdS-black brane on $\mathcal{B}$ are decoupled ($i.e.$ imposing the reflecting boundary conditions on the boundary of $\mathcal{B}$); at $\mathfrak{t}$=0, the CFT reservoir and the AdS-black brane are coupled and the black brane on $\mathcal{B}$ start evaporating but fixing the temperature. At Page time, quantum extremal surfaces appear and the entanglement entropy stops increasing.

\subsection{Induced Gravity on Planck Brane}
Our goal is to derive a way to embed the $(n+1)-$dimensional Kerr-Newman black brane into the $(n+2)-$dimensional asymptotically Kerr-Newman black brane. In this subsection, we will analytically show how to embed the $AdS_{n+1}$ into AdS$_{n+2}$, that is, the Hartle-Hawking state on pure AdS with zero temperature auxiliary reservoir. In the general case, the $(n+1)-$dimensional AdS black brane on $\mathcal{B}$ with finite temperature thermal bath, the bulk geometry $\mathcal{M}$ is difficult to write down analytically, but~\cite{Almheiri:202007} provides a nice way to numerically construct the bulk geometry by using the Deturck trick. Especially, the bulk geometry become exact AdS black brane while we consider weak tension limit on $\mathcal{B}$.

Start from the $AdS$ foliated metric
\begin{align}\label{foliated-metric}
    ds^{2}_{AdS_{n+2}}&=\frac{L^2}{\sin^2{\theta}}\left(d\theta^2+\frac{ds^{2}_{AdS_{n+1}}}{L^2}\right)\\
    &=\frac{L^2}{z^2}\left(-dt^2+dz^2+\ell^2 d\phi^2+d\sigma^2_{n-1}\right)\,.\label{pcradsbulk}
\end{align}
Later we will consider the cylindrical rotating black brane as the bulk, the length $\ell$ is the radius of the cylinder and the coordinate $\phi$ is compactified as $\phi \sim \phi + 2\pi$. To simplify our notation, we let $\ell$ to be $1$. Let's say the brane is located at $\theta=\theta_\mathcal{B}$, then the induced metric on $\mathcal{B}$ is
\be\label{ads-(n+1)}
ds^{2}_{\mathcal{B}}=\frac{L_\mathcal{B}^2}{\omega^2}\left(-dt^2+d\omega^2+d\sigma^2_{n-1}\right)\,.
\ee
Note that we have the following relations:
\begin{align}\label{zeta-omega}
    z=\omega\sin \theta,~~~ \phi=\omega\cos \theta\,.
\end{align}
By the Israel junction condition (\ref{eom-on-b}), we are able to compute the tension on  $\mathcal{B}$
\be\label{tension}
\mathcal{T}_\mathcal{B}=\frac{n}{L}\cos{\theta_\mathcal{B}}\,.
\ee
From (\ref{foliated-metric}), (\ref{ads-(n+1)}) and (\ref{tension}), it is easy to read off that the length scale $L_\mathcal{B}$ on brane is
\be
\frac{1}{L_\mathcal{B}^2}=\frac{\sin^2{\theta_\mathcal{B}}}{L^2}=\frac{1}{L^2}\left(1-\frac{L^2\mathcal{T}_\mathcal{B}}{n^2}\right)\,.
\ee
In general, the bulk manifold $\mathcal{M}$ is not an AdS black hole \cite{Almheiri:202007} because the black hole on the Planck brane will break the translational symmetry in $\phi$ direction. The following we analytically calculate the entanglement entropy in the case that the Plank brane with weak tension limit ($\theta_{\mathcal{B}} \rightarrow \pi/2$), that means its backreaction to the bulk geometry is negligible.
\section{Time Evolution of Wormhole Area}\label{sec3}
In the no island phase, the RT surface $\Gamma$ is the wormhole path (the green lines in Fig.~\ref{phae}). This quantity is also related to the time evolution of holographic complexity~\cite{Carmi:2018} for computational complexity=volume (CV) conjectures. In this section, we are going to explore the time evolution of $\mathcal{A}_{\Gamma,\mathcal{I}=\{\emptyset\}}$.
\subsection{Dynamics of Wormhole}
The line element of a cylindrical Kerr-Newman black brane becomes~\cite{Awad:2002cz}
\begin{equation} \label{metric}
ds^{2} = \frac{L^2}{z^2} \left[ - h(z) (\Theta \, dt - a \, d\phi)^2 + \frac{dz^2}{h(z)} + (\Theta \, d\phi - a \, dt)^2 + d\sigma_{n-1}^2 \right],
\end{equation}
Let's rewrite the Kerr-Newman metric (\ref{metric}) in Eddington–Finkelstein coordinates
\be
ds^2_{\mathcal{M}}=\frac{L^2}{z^2}\Big[-dv^2 \left(1-\Theta ^2 g(z)\right)-2 dv dz\Theta+2 a d\psi (dz-dv \Theta  g(z))+ d\psi^2 \left(1+a^2g(z)\right)+d\sigma_{n-1}^2\Big]
\ee
where 
\be \label{tildevtz}
dv=dt-\frac{\Theta }{h(z)}dz~~~d\psi=d\phi-\frac{a }{h(z)}dz\,.
\ee
It is convenient to use the radial direction\footnote{Here we assumed $\lambda \propto r^{\alpha>0}=\frac{1}{z^{\alpha>0}}$} $\lambda$ to parameterized the extremal surface $\Gamma$
\be\label{area-no-island}
\mathcal{A}_{\Gamma,\mathcal{I}=\{\emptyset\}}=L^n\mathcal{A}_{\partial \Upsilon}\int_{\lambda_{m}}^{\lambda_{UV}}\mathcal{L}(\dot{v},\dot{z},v,z;\lambda)\,d\lambda
\ee
with
\begin{align}
\mathcal{L}(\dot{v},\dot{z},v,z;\lambda)=\frac{1}{z^{n}}\sqrt{-\dot{v}^2 \left(1-\Theta^2 g\left(z\right)\right)-2 \Theta  \dot{v} \dot{z}-\frac{2 a^2\dot{z}}{h\left(z\right)}\left(\dot{z}-\Theta  g\left(z\right) \dot{v}\right)+\frac{a^2 \dot{z}^2}{h\left(z\right)^2}\left(a^2 g\left(z\right)+1\right)}
\end{align}
where $z(\lambda_{UV})=z_{UV}$ is the UV cutoff and $\lambda_m$ is the turning point of $z$, that is , $\dot{z}(\lambda_m)=0$. Since Lagrangian does not explicitly depend on $\lambda$ in ($\ref{area-no-island}$), we can choose the appropriate $\lambda$ so that
\be\label{guage}
\mathcal{L}(\dot{v},\dot{z},v,z;\lambda)=L^{n-1}\,.
\ee
By varying the variables $v$ and $z$ and applying (\ref{guage}), we obtained the equation of motion
\begin{align}\label{eom1}
    &\partial_\lambda\left[\frac{\left(1-\Theta ^2 g\left(z\right)\right)}{h\left(z\right) z^{2 n}} \left(\dot{v} h\left(z\right)+\Theta\dot{z}\right)\right]=0\\
    &\partial_\lambda\left[\frac{1}{z^{2n}}\left(\Theta\dot{v}\frac{\left(1-\Theta ^2 g\left(z\right)\right)}{h\left(z\right)}+a^2\dot{z}\frac{\left(1-(1+\Theta^2) g\left(z\right)\right)}{h\left(z\right)^2}\right)\right]=\frac{nL^{2n-2}}{z}-\frac{\partial_{z} g\left(z\right)}{2 z^{2n}}\tilde{K}\label{eom2}
\end{align}
with $\tilde{K}\equiv \Theta^2\dot{v}^2+2 a^2 \Theta \frac{\dot{v} \dot{z}}{h\left(z\right)^2}+\frac{a^2 \dot{z}^2}{h\left(z\right)^3}\left(a^2+\left(\Theta ^2+1\right) g\left(z\right)\right)$. Observing (\ref{eom1}), there is a constant of motion 
\be\label{e}
E=\frac{\left(1-\Theta ^2 g(z)\right)}{h(z) z^{2 n}} \left(\dot{v} h(z)+\Theta\dot{z}\right),~~~h(z)=1-g(z)
\ee
due to $\partial_{v} h\left(z\right)=0$ in the left-hand side of (\ref{eom1}). Via (\ref{e}), we get the relation between $v(\lambda)$ and $z(\lambda)$ 
\be\label{vot}
\dot{v}=\frac{E z^{2n}-\frac{(1-\Theta^2 g(z))}{h(z)}\Theta\dot{z}}{(1-\Theta^2 g(z))}\,.
\ee
Plugging (\ref{vot}) into (\ref{guage}), we obtained the equation of motion for $z(\lambda)$ 
\be\label{eom-z0t}
\dot{z}=-L^{n-1}z^{n}h(z)\sqrt{\frac{1-\Theta^2 g(z)+E^2L^{2-2n}z^{2n}}{(1-\Theta^2 g(z))h(z)}}\,.
\ee
If we know the value $E$, form (\ref{eom-z0t}), we can determine the turning point by
\be\label{e-zm}
1-\Theta^2g(z_{m})+L^{2-2n}E^2 z_{m}^{2(n+1)}=0,~~~\text{and}~~~z_{m}\equiv  z(\lambda_m)\,.
\ee
It is convenient to measure this thermalization process by boundary time $v(\lambda_{UV})\equiv\mathfrak{t}$, then we have $\mathfrak{t}=v(\lambda_{m})+\int_{z_{UV}}^{z_m} \left(-\frac{\dot{v}}{\dot{z}}\right)\,dz\,$. Using the relation in (\ref{tildevtz}), we get $v(\lambda_{m})=\int_{0}^{z_{m}} \frac{\Theta dz}{h(z)}$. Here we conclude
\begin{align}
    \mathfrak{t}=\int_{0}^{z_{m}} \frac{\Theta}{h(z)}dz+\int_{z_{UV}}^{z_{m}}\frac{1}{h(z)}\left(\frac{E z^{n}}{(1-\Theta^2 g(z))L^{n-1}}\sqrt{\frac{(1-\Theta^2 g(z))h(z)}{1-\Theta^2 g(z)+E^2L^{2-2n}z^{2n}}}+\Theta\right)\,dz\,.\label{int-time}
\end{align}
Applying (\ref{area-no-island}), (\ref{guage}) and (\ref{eom-z0t}), we can formulate
\be\label{int-area-no-island}
\mathcal{A}_{\Gamma,\mathcal{I}=\{\emptyset\}}= L^n\mathcal{A}_{\partial \Upsilon}\int_{z_{UV}}^{z_{m}}\frac{1}{z^{n}h(z)}\sqrt{\frac{(1-\Theta^2 g(z))h(z)}{1-\Theta^2 g(z)+E^2L^{2-2n}z^{2n}}}\,dz\,.
\ee
From (\ref{int-time}) and (\ref{int-area-no-island}), we obtain the derivative wormhole area with respect to the boundary time
\be\label{dadt-no-island}
\frac{d\mathcal{A}_{\Gamma,\mathcal{I}=\{\emptyset\}}}{d\mathfrak{t}}= L^n \mathcal{A}_{\partial \Upsilon} \frac{\sqrt{\abs{1-\Theta^2g(z_{m})}}}{z_{m}^{n}}=\mathcal{A}_{\partial \Upsilon}\abs{E}L\,.
\ee
For the second equality in (\ref{dadt-no-island}), we have used (\ref{e-zm}). Here we see the additional rotation correction $\Theta\in[1,\infty)$ in (\ref{dadt-no-island}). As rotating parameter $a$ increase, so does the entanglement entropy. Note that the result (\ref{dadt-no-island}) return to the charged black hole case \cite{Ling:202010,Carmi:2018} as the rotating parameter $a$ vanishes ($i.e. \Theta=1$). We emphasize that the entanglement growth is independent of $\phi$ coordinate owing to the translational symmetry of spacetime, so that the ambiguity of how the increase of the radiation regions $\Upsilon$ is not important. The reason why area time-dependent is that wormhole itself is time-dependent.
\subsection{Late Times}
We suppose that the evaporation is tremendously slow such that $\mathfrak{t}_P\rightarrow \infty$. In the late times' regime, the entanglement grows linearly due to there exist the critical surface $z_{mc}$ that a great part of the extremal surface runs along this critical surface $z_{mc}$~\cite{Hartman:2013}
\be\label{linear}
\Delta\mathcal{A}_{\Gamma,\mathcal{I}=\{\emptyset\}}\simeq L^n \mathcal{A}_{\partial \Upsilon}v_{E}\mathfrak{t}
\ee
where the entanglement velocity
\be\label{ve}
v_{E}= \frac{\sqrt{\abs{1-\Theta^2g(z_{mc})}}}{z_{mc}^{n}}
\ee
can be computed by finding the critical surface $z_{mc}$
\begin{align}
    \label{critical}
\Eval{\frac{d}{dz_{m}}\left(\frac{\sqrt{\abs{1-\Theta^2g(z_{m})}}}{z_{m}^{n+1}}\right)}{z_{m}=z_{mc}}{}=0~~~\Rightarrow ~~~ z_{mc}=\left(\frac{2n}{m (n-1)\Theta^2}\right)^{\frac{1}{n+1}}\,.
\end{align}
Specifically, $v_{E}=\sqrt{\frac{n+1}{n-1}}\left(\frac{2n}{m (n-1)\Theta^2}\right)^{-\frac{n}{n+1}}$ in the Kerr case ($q=0$). We conclude that if the characteristic  of Kerr-Newman black hole ($\mathcal{E}, \mathcal{Q}, \mathcal{J}$) and the spacetime dimension $n$ are known, the entanglement velocity can be determined.
For Kerr black brane, it is easy to see $v_{E}=0$ at zero temperature. For RN black brane, \cite{Ling:202010} has shown some numerical results which indicate the entanglement velocity vanishes $v_{E}=0$ in the extremal case.

\section{Saturation and Quantum Extremal Island}\label{sec4}
\begin{figure}
\centering
\includegraphics[width=0.7\textwidth]{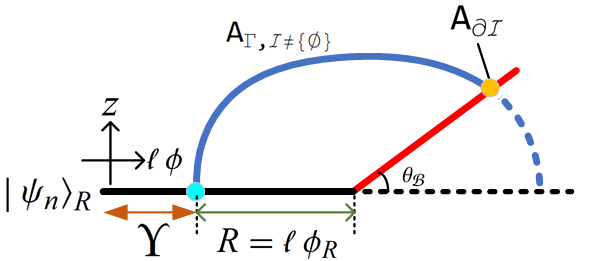}
\caption{When the islands appear, the RT surface is equal to the minimal surface defect given the strip in the boundary.}
\end{figure}
We have seen that the wormhole area will keep growing linearly (\ref{linear}) in late times until the island appears. There is an ambiguity that we do not know how big the radiation region is. Having said that, we can find the relation between the size of radiation regions and the entanglement entropy of Hawking radiation. 

From \cite{Miao:2020}, the relation between the Newton's constant in bulk $\mathcal{M}$ and brane $\mathcal{B}$ is
\be
G^{(\mathcal{M})}_N=G^{(\mathcal{B})}_N L\int_{\pi/2}^{\theta}\csc^{n-1}(y)\,dy\,.
\ee
If we take $\theta_{\mathcal{B}} \rightarrow \pi/2$ limit, the Newton's constant in bulk $G^{(\mathcal{M})}_N$ is much smaller than the Newton's constant in brane $G^{(\mathcal{B})}_N$, which means the entanglement entropy in the island phase is dominated by the quantum part of (\ref{island-rule}) and the $\mathcal{A}_{\partial \mathcal{I}}/(4G^{(\mathcal{B})}_N)$ in (\ref{evap-ee}) is neglectable. Moreover, in a such limit, the von Neumann entropy of Hawking radiation is approximately identical to the HEE for the strip with width $2R$, $i.e.$ $\mathcal{A}_{\Gamma,\mathcal{I}\neq\{\emptyset\}}\simeq 2\mathcal{A}_{strip}(\phi_{R})$. In this section, we just conclude that when island appears, the entanglement entropy of Hawking radiation in the small $R$ limit \footnote{The intersection between the RT surface $\Gamma$ and the Plank brane $\mathcal{B}$ is far from the event horizon.} is~\cite{pcSun:2021}
\be\label{ee-island-s}
\Delta\mathcal{S}_\Upsilon\equiv\mathcal{S}_\Upsilon-\mathcal{S}_{\Upsilon(0)}\simeq m\frac{L^{n}\mathcal{A}_{\partial \Upsilon}}{4 G^{(\mathcal{M})}_N}\left(\frac{n}{2}+a^2 \right) \frac{\tilde{R}_1}{\tilde{R}^2_0}\phi_{R}^{2}
\ee
in which $\tilde{R}_0 \equiv \frac{\sqrt{\pi} \, \Gamma\left( \frac12 + \frac1{2 n} \right)}{\Gamma\left( \frac{1}{2 n} \right)}$, $\tilde{R}_1 \equiv \frac{\sqrt{\pi} \, \Gamma\left( \frac1{n} \right)}{2 n (n + 2)\Gamma\left( \frac12 + \frac1{n} \right)}$ and $\mathcal{S}_{\Upsilon(0)}$ is the entanglement entropy in vacuum. Also, in the large $R$ limit\footnote{The intersection between the RT surface $\Gamma$ and the Plank brane $\mathcal{B}$ is very close to the event horizon.}, the leading behavior of entanglement entropy is   
\be\label{ee-island-l}
\Delta\mathcal{S}_\Upsilon\simeq\frac{L^{n}\mathcal{A}_{\partial \Upsilon}}{4 G^{(\mathcal{M})}_N}\frac{\Theta \,\phi_{R}}{z_h^n}.
\ee
We see that, in the small $R$ limit, the charge parameter $q$ does not appear in the leading behavior (\ref{ee-island-s}); in the large $R$ limit, the leading behavior of entanglement entropy is equal to the area of horizon divided by $4G^{(\mathcal{M})}_N$, which is Bekenstein-Hawking entropy. Note that, in both small and large $R$ limits, the entanglement entropy increase as rotating parameter $a$ increases.

\section{Conclusions and Outlooks}\label{sec5}
We have analytically investigated the Page-like curve for the Kerr-Newman black brane $(k=0)$ at the semi-classical level by the holographic approach. Although to study the more realistic problem, we should consider the black hole with cosmological constant $\Lambda=0$ and the topology of the boundary (located at $r=\infty$) is the sphere $(k=1)$, at least, there is a chance to discover pedagogically some hints about the black information paradox. Besides, we will focus the discussion on how the rotating parameter affects the von Neumann entropy of Hawking radiation.

Before Page time, from (\ref{dadt-no-island}), we see the rate of entanglement growth is proportional to the rotating parameter and mass parameter. In particular, in the late times, there is a great part of the wormhole along the critical surface, which leads the linear growth. There are similar properties occurs that occur in the situation that the entanglement entropy during the thermalization is discussed in~\cite{Liu:2013iza,Liu:2013qca}. We discover that the entanglement velocity $v_E$ in (\ref{ve}) is only determined by the emblackening factor $h(z)$ and the rotating parameter $a$, which characterizes the feature of the equilibrium state of the black brane. Notice that when the rotating parameter $a$ increases, the entanglement velocity $v_E$ speeds up. That means the faster the black brane rotates, the entangled Hawking particles to be radiated faster in late times.

\begin{acknowledgments}
I would like to thank Professor Chiang-Mei Chen for the valuable discussion. The work was supported in part by the Ministry of Science and Technology, Taiwan.
\end{acknowledgments}

\end{document}